\documentclass{article}
\usepackage{latexsym,aaspp4,times,flushrt,epsf}
\def\etal    {et~al.} 
\def\eg      {e.g.}

\def\ie      {i.e.}
\def\R       {{\bf\it R}}
\def\rms     {r.m.s.}
\def\fdg     {\hbox{$.\!\!^\circ$}}
\def\kms     {\ifmmode{\mbox{km~s}^{-1}}\else{km~s$^{-1}$}\fi} 
\def\dagger  {\ifmmode{^\dag}\else{$^\dag$}\fi} 
\def\onedrms {\ifmmode{\sigma_{1D}}\else{$\sigma_{1D}$}\fi}
\def\rqrt    {\ifmmode{r^{1/4}}\else{$r^{1/4}$}\fi}
\def\avgfp   {\ifmmode{\overline{\mbox{FP}}}\else{$\overline{\mbox{FP}}$}\fi}
\def\fpscat  {\ifmmode{\Sigma}\else{$\Sigma$}\fi}
\def\muscat  {\ifmmode{\langle\Sigma\rangle}\else{$\langle\Sigma\rangle$}\fi}
\def\muvar   {\ifmmode{\langle\Sigma^2\rangle}\else{$\langle\Sigma^2\rangle$}\fi}
\def\logsigma{\ifmmode{\mbox{log}(\sigma)}\else{log$(\sigma)$}\fi}
\def\reff    {\ifmmode{r_{e}}\else{$r_{e}$}\fi}
\def\mueff   {\ifmmode{\langle\mu\rangle_{e}}\else{$\langle\mu\rangle_{e}$}\fi}
\def\mle     {\ifmmode{{\mathcal MLE}}\else{${\mathcal MLE}$}\fi}
\def\mlf     {\ifmmode{{\mathcal L}}\else{${\mathcal L}$}\fi}

\lefthead{FUNDAMENTAL PLANE AND PECULIAR VELOCITIES}
\righthead{GIBBONS, FRUCHTER, \& BOTHUN}

\begin{document}
\doublespace

\title{Deviations from the Fundamental Plane and the Peculiar
       Velocities of Clusters}

\author{R. A. Gibbons\altaffilmark{1} and A. S. Fruchter\altaffilmark{2}}
\affil{Space Telescope Science Institute, 
       3700 San Martin Drive, 
       Baltimore, MD 21218;}
\affil{gibbons@stsci.edu; fruchter@stsci.edu}

\author{G. D. Bothun\altaffilmark{2}}
\affil{Department of Physics, University of Oregon, 
       120 Willamette Hall, 
       Eugene, OR 97403;}
\affil{nuts@moo2.uoregon.edu}

\altaffiltext{1}{Department of Astronomy, University of Maryland, 
                 College Park, MD 20742}
\altaffiltext{2}{Visiting Astronomer, Kitt Peak National Observatory, 
                 National Optical Astronomy Observatories, which is 
                 operated by the Association of Universities for Research 
                 in Astronomy, Inc. (AURA) under cooperative agreement with 
                 the National Science Foundation.}

\begin{abstract} 
We have fit the Fundamental Plane of Ellipticals (FP) to over 400 early-type
galaxies in 20 nearby clusters (cz~$\sim4000-11000$~\kms), using our
own photometry and spectroscopy as well as measurements culled from the
literature. We find that the quality-of-fit, \rms[\logsigma], 
to the average Fundamental Plane (\avgfp) varies substantially among these
clusters. A statistically significant gap in \rms[\logsigma] roughly separates
the clusters which fit \avgfp\ well from those that do not.

In addition, these two groups of clusters show distinctly different behavior 
in their peculiar velocity (PV) distributions. Assuming galaxies are drawn 
from a single underlying population, cluster PV should not be correlated with 
\rms[\logsigma]. Instead, the clusters with below average scatter
display no motion with respect to the cosmic microwave background
(CMB) within our measurement errors ($\sim250$~\kms), while clusters in the 
poor-fit group typically show large PVs.
Furthermore, we find that all X-ray bright clusters in our sample fit 
the \avgfp\ well, suggesting that early-type galaxies in the most massive, 
virialized clusters form a more uniform population than do cluster ellipticals 
as a whole, and that these clusters participate in a quiet Hubble flow.

\end{abstract}


\section{Introduction}

\subsection{\it The Fundamental Plane of Ellipticals}
The high correlation of the structural and kinematic properties of ellipticals
suggests that these galaxies formed by similar processes and are largely
virialized. Assuming structural symmetry, isotropic velocities, 
$\langle$$v^2\rangle=\sigma^2$, a constant mass-to-light ratio (M/L), 
and structural homology, a tight correlation is expected as a natural 
result of virialization :
\begin{equation}\label{eq:fpvirial} 
\sigma^2 \propto \langle SB\rangle r ,
\end{equation} 
where $\sigma^2=$ the velocity dispersion within the galaxies, $r=$ 
a fiducial radius, and $\langle$$SB\rangle$ the average surface brightness 
within $r$. In log space this expression relating basic physical properties
collapses to a plane and, hence, is called the Fundamental Plane of Ellipticals
(Djorgovski \& Davis 1987; Lynden-Bell \etal\ 1988; 
Lucey, Bower, \& Ellis 1991). In terms of observables the FP becomes
\begin{equation}\label{eq:fpobservables}
\mbox{log}(\reff) = \alpha\;\mbox{log}(\sigma) + \beta\;\mueff + Const ,
\end{equation}
with $\alpha=2$, $\beta=0.4$ representing the virial plane. Standard units of 
measure are \kms\ for the central velocity dispersion, $\sigma$, arcseconds 
for the \rqrt$-$law half-light radius, \reff (effective radius), and 
magnitudes per arcsec$^2$ for the average surface brightness within that 
radius, \mueff (effective surface brightness).

Empirically, ellipticals occupy a narrow range within this three-dimensional
parameter space indicating they are a homologous family. S0 galaxies also lie 
on the FP due to the fact that, while errors in the estimated \reff\ and 
\mueff\ are made by fitting \rqrt$-$law profiles to galaxies with disk 
components, the errors cancel such that the FP quantity, 
``$mbox{log}(\reff) - \beta\;\mueff$'', is left remarkably unaffected 
(Saglia \etal\ 1997;
Scodeggio \etal\ 1998; Kelson \etal\ 1999). The observed FP, however, is 
misaligned with the virial prediction. Ellipticals are 
better described by, $\alpha\sim1.4$ and $\beta\sim0.3$. This tilt of the FP 
implies the assumptions used in Equation~\ref{eq:fpvirial} are not quite true.
Nonetheless, the uniformity of ellipticals makes these galaxies important 
standard candles since the observed \reff\ provides a direct indication of 
distance (Dressler 1987; Dressler \etal\ 1987; Lucey \& Carter 1988). The 
small observed scatter about the FP translates to $\sim19\%$ error in the 
distance to individual galaxies 
(J{\o}rgensen, Franx, \& Kj{\ae}rgaard 1996; Hudson \etal\ 1997; present work).
The FP is thus comparable as a distance indicator to the Tully-Fisher (TF) 
relation applied to spirals (de~Carvalho \& Djorgovski 1989). 

\subsection{\it Motivation \& Goal for this Study}

The FP, therefore, should be a powerful tool for measuring relative cluster
distances as well as deviations from the Hubble flow. However, surveys of large
scale motions using the FP and TF have found significantly different cosmic 
flows.  For example, the recent FP ``Streaming Motions of Abell Clusters'' 
survey (SMAC) (Hudson \etal\ 1999) reveals a large bulk peculiar motion on the 
galaxy cluster scale.  Yet, while SMAC agrees with the 15 cluster TF survey 
(LP10K) of Willick \etal\ 1999, most TF surveys have shown little peculiar 
motion on these scales. The large and somewhat deeper TF cluster survey of 
Dale \etal\ 1999a limits the motion to less than $200~\kms$, agreeing with TF 
surveys going back to the Aaronson \etal\ 1986 TF survey. To further confuse 
the issue, Lauer \& Postman 1994, using a novel method employing brightest 
cluster galaxies in 119 clusters, detected a large bulk flow of the 
``Abell Cluster Inertial Frame'' (ACIF). But the direction is in conflict with 
the results of the aforementioned surveys, none of which have the large sky 
coverage of the ACIF survey.

As can be seen from the compilation of previous measurements 
(Table~\ref{tab:surveys}), the results of these surveys are clearly discrepant
or their errors have been underestimated. Indeed, some of the possible sources
of systematic errors, as summarized by Jacoby \etal\ (1992), are still
not well understood. Moreover, as emphasized by 
Strauss \etal\ 1995, if flows on this large a scale exist, then they can not be
accommodated by any present cosmological model. These models assume the 
average random motion of clusters atop the Hubble flow is driven by gravity 
and the underlying mass distribution and thereby provides a constraint on 
the density parameter, $\Omega$\
(Borgani \etal\ 1997; Watkins 1997; Bahcall \& Fan 1998).

To address this problem, we examine the FP and rigorously test the assumption 
that the FP is invariant from cluster to cluster. If this assumption is 
violated, then previous surveys, which have assumed a uniform fundamental
plane, could contain significant systematic errors. We have compiled a sample 
of over 400 early-type galaxy in 20 clusters. Measured distances to clusters 
should be more accurate than to individual galaxies by the root number of 
galaxies observed per cluster, and thus typical distance errors to individual 
clusters should be of order $\sim5\%$. Hence, flows on cluster scales can in 
principle be accurately determined. Furthermore, this sample allows us to 
investigate the effects of possible systematic errors on a cluster by cluster 
basis. The rest of this paper is organized as follows: in \S2 we briefly 
describe our data as a thorough presentation will be the subject of a 
subsequent paper; we discuss our FP fitting procedure in \S3; results and 
statistical analysis are presented in \S4; we finish with further discussion 
and implications of our result in \S5; and summarize our conclusions in \S6.

\section{\label{Data}The Data}
Our observations have yielded useful spectroscopy and photometry of 132 E+S0
galaxies in 8 Abell clusters. Spectra have been obtained with the Nessie multi 
fiber instrument on the KPNO Mayall 4m, and imaging with a large format 
($2048^2$) Tek CCD on the KPNO 0.9m. Redshifts and central velocity 
dispersions are measured by matching the galaxy spectra with G and K class 
stellar templates, using the Fourier-quotient technique (Sargent \etal\ 1977;
Tonry \& Davis 1979) 
adapted for use in IRAF by Kriss (1992). Aperture corrections are applied to 
the velocity dispersions, due to the fact that the fibers, which are of fixed
diameter, sample larger portions of galaxies at larger distances, 
systematically lowering measured velocity dispersions. For model ellipticals,
J{\o}rgensen, Franx, \& Kj{\ae}rgaard 1995 find the velocity 
dispersion measured within an aperture varies with aperture size as a power 
law of index of $-0.04$.
The applied aperture correction to measured velocity dispersions is then
calculated using cluster redshift as a first order estimate of distance and 
the angular~diameter$-$distance relation with $q_0=0.5$.

Photometry has been corrected for atmospheric extinction, Galactic 
absorption (Burstein \& Heiles 1984), $(1+z)^4$ cosmological dimming,
and {\it k}-correction. The {\it k}-correction to flux scales as $(1+1.089z)$, 
calculated for an average elliptical galaxy spectrum 
(Coleman, Wu, \& Weedman 1980) through our CCD and the \R\ filter. We work in 
\R\ band to help minimize the effects of cluster differences in age and 
metallicity which can cause shifts in the FP at shorter wavelengths
(Guzm\'{a}n \& Lucey 1993; Gregg 1995). Effective radii are calculated by 
fitting an \rqrt$-$law to isophotes, which have been obtained using the 
ELLIPSE task in IRAF. A grid of seeing corrections to \reff\ and \mueff\ is 
calculated from models convolved with the PSFs of the images.

Merging our data with recently published observations, increases the number of 
galaxies to 428 in 20 clusters. Table~\ref{tab:sample} lists the literature 
sets as well as abbreviations used hereafter. Velocity dispersions, effective 
radii, and surface brightnesses have been normalized to the SLHS97 system by 
minimizing residuals of each of these parameters for 56 galaxies in four 
clusters common to both samples. The details of our observations and this 
conversion including seeing and aperture corrections will be described in 
detail in Gibbons, Fruchter, \& Bothun (2000c). For some clusters, \eg\ A400, 
our sample size is limited because we exclude objects with strong emission 
lines and evidence of a strong disk component. We retain only galaxies which 
are clearly cluster members and which have excellent spectroscopic S/N. 
Average galaxy FP errors for clusters are comparable between the three data 
sets except for A400, whose spectra are of lower quality than the other 
clusters. However, excluding this cluster from our analysis does not alter 
our results.

\section{\label{Fit}Fitting to All Data Simultaneously}
Once the corrections described in \S2 are applied to the measured $\sigma$'s 
and \mueff's, shifts in the FP should be due purely to the change in apparent 
galaxy size with distance. The \avgfp\ is then fit to all the galaxy data
simultaneously by letting cluster distance be a free parameter. Specifically, 
we solve a set of equations with 428 galaxies in 20 clusters,
\begin{equation}\label{eq:logr} 
\mbox{log}(\reff) = \alpha\;\logsigma + \beta\;\mueff + \gamma_{i}\;,\; 
i = 1,20
\end{equation} 
where shifts in the intercepts, $\Delta\gamma_{i}$, reflect the offsets in 
log(\reff), which translate into the relative distances between clusters.

Because one is seeking high precision in relative distances it may
seem most prudent to minimize the residuals about the FP in the direction of 
the distance dependent parameter, log(\reff). This particular projection also 
yields the smallest observed scatter due to the strong correlation between 
\reff\ and \mueff. However, Lucey, Bower, \& Ellis (1991) as well as 
J{\o}rgensen, Franx, \& Kj{\ae}rgaard (1996) have recognized that minimizing 
the residuals of log(\reff) will introduce a bias into the fit because the 
errors in log(\reff) and \mueff\ are correlated, since \mueff\ is a function of
\reff\ (see also Akritas \& Bershady 1996). 

Therefore, like SLHS97, we isolate the velocity dispersion, $\sigma$, as the 
only parameter with independent errors. Explicitly, we write the FP solely as 
a function of \reff,
\begin{equation}\label{eq:logsigma} 
\logsigma=f(\reff)=(\mbox{log}(\reff)-\beta\;\mueff-\gamma_{i})/\alpha\;,\; 
i = 1,20 
\end{equation} 
and solve for the coefficients, which minimize the absolute residuals,
\begin{equation}\label{eq:minsum} 
\sum_{i=1,20}\sum_{j=1,n_i}|\logsigma_j-f_{ij}(\reff)|
\end{equation}
where $n_i=$ the number of galaxies per cluster. When the absolute residuals 
are minimized, galaxies which lie far off the main relation have effectively 
less weight than they would under least squares. Our use of $\sigma$ 
as the independent variable avoids the strong bias caused by the correlated 
errors in \reff\ and \mueff\ when \reff\ is instead treated as the independent 
variable.

We solve this system of equations allowing 22 free parameters: 20 cluster 
intercepts, $\gamma_{i}$'s, plus common $\alpha$ and $\beta$. In this way, 
we find the average fundamental plane, \avgfp, which best fits the entire 
sample of galaxies, further reducing the influence of anomalous galaxies. To 
ensure the best solution within a parameter space of such complexity (22 
dimensions) requires an efficient searching algorithm. We adopt simulated 
annealing, because it efficiently converges towards the global minimum 
while avoiding becoming trapped in local minima (Metropolis \etal\ 1953;
Kirkpatrick, Gelatt, \& Vecchi 1983; Vanderbilt \& Louie 1984). The convergence
is quite rapid and most of the computing time is spent sampling the region 
near best fit.

\section{\label{Results}Results}

\subsection{\it The FP Derived from the Present Sample}
\begin{figure}
\plotone{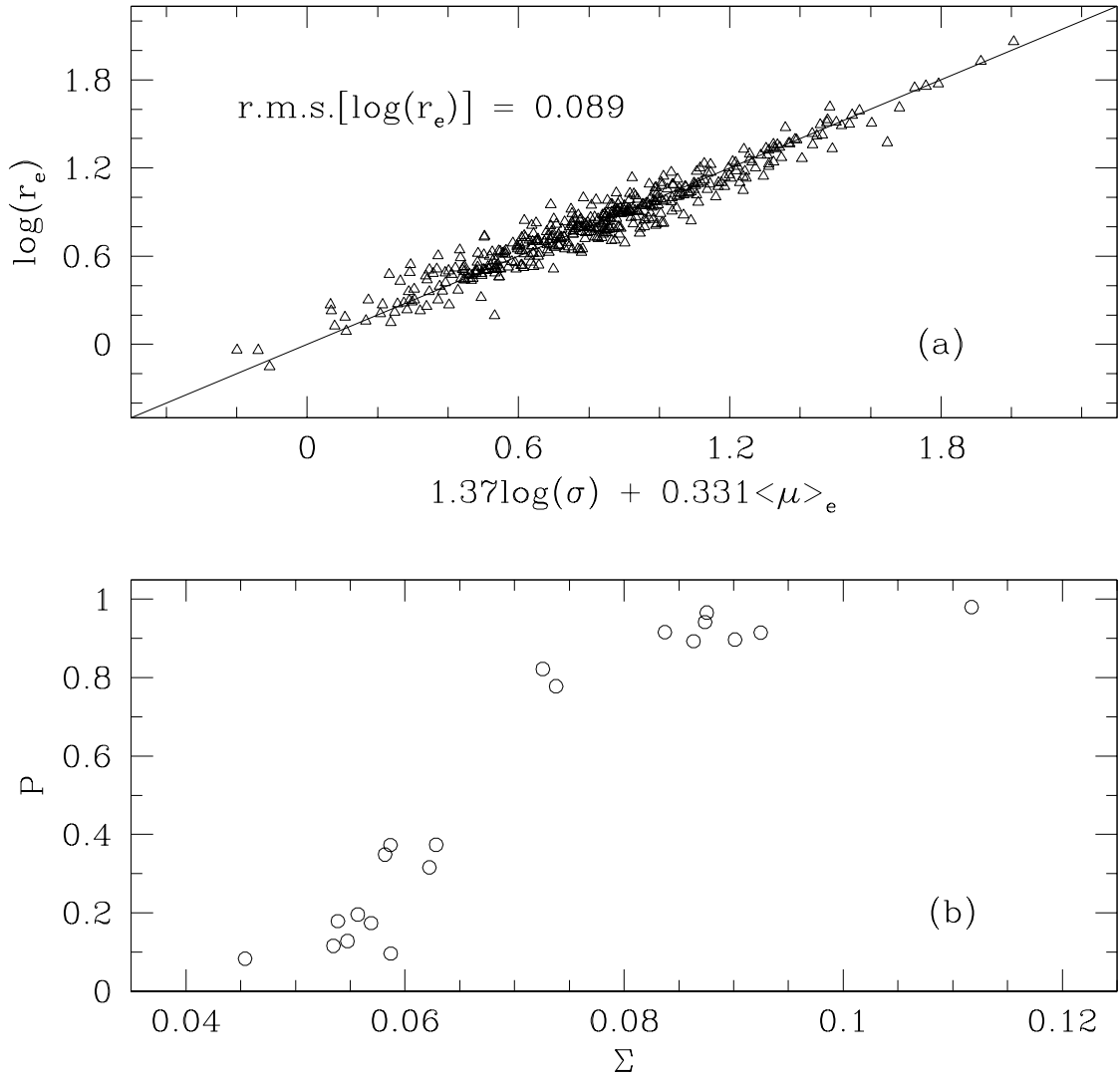}
\caption{\label{fig:fp20}(a) The \avgfp\ for the entire sample of 428 
         galaxies. This projection shows the scatter about log(\reff). 
         (b) The significance of $\chi^2$ about \avgfp\ for each cluster vs. 
         FP scatter. The gap seen in probability space is highly significant,
         making it unlikely that all clusters have been drawn from a single 
         population.}
\end{figure}

The best-fit \avgfp, $\alpha=1.37\pm$$0.04$, $\beta=0.331\pm$$0.004$, is shown
in Figure~\ref{fig:fp20}a and is in agreement with that found 
by Hudson \etal\ (1997). We estimate the errors on $\alpha$ and $\beta$ by 
fixing $\gamma_{i}$, varying $\alpha$ and $\beta$, and constructing maps of 
chi-square versus $\alpha$ and $\beta$. Our chi-square assumes the average 
total FP scatter (intrinsic plus measurement error) is well represented by
the measured average \rms\ scatter in \logsigma\ about \avgfp, \muscat. Then
\begin{equation}\label{eq:chifp}
\chi^2=\sum_{j=1}^{n}[\mbox{log}(\sigma_j)-\mbox{\avgfp}]^2/
\langle\Sigma^2\rangle .
\end{equation}
The one sigma errors, corresponding to $\Delta\chi^2=1$, are read directly from
these curves. For our fit, the average \rms\ scatter in log[$\sigma$],
$\muscat=0.065$, is equivalent to a $20\%$ error in distance to individual 
galaxies. Assuming the average scatter applies to all clusters, errors in 
{\it cluster} distances range from about $2-8\%$, depending on the number of
galaxies fit per cluster. 
  
However, 9 of the 20 clusters, about half the total sample, show significantly 
more scatter than the other 11 (hereafter \fpscat\ denotes FP \rms\ scatter). 
When these clusters are fit separately, they also fail to define a tight FP.
We must therefore investigate the significance of the larger scatter: \ie, 
whether or not the high-\fpscat\ fits are truly poor. The degree to which the 
\avgfp\ describes the individual clusters can be seen in 
Figure~\ref{fig:fp20}b. Assuming galaxies in a cluster are chosen randomly 
from a single population, we calculate for each cluster the significance, or 
probability, $P$, of a chi-square this large about the \avgfp\ and plot this 
value against \fpscat. Immediately evident is a break in $P$ coincident with a
separation in \fpscat\ about \muscat.
While the clusters should uniformly fill probability
space, a gap this large anywhere in $P$ has less than a $0.1\%$ chance of 
occurring (Similarly, along the \fpscat\ axis, one would expect the points to 
be clustered about \muscat\ rather than showing the deficit that is observed).
However, as this estimate of probability relies on the assumption that the
FP is gaussian, we have also estimated the
probability of this gap by simulating many trial sets of clusters from the 
total sample of observed galaxies and calculating the distribution of the 
maximum gap. Doing so again implies the observed situation is improbable, with 
an expected occurrence of a gap this size of $0.2\%$.
This strongly suggests that our sample has not been drawn from a single galaxy 
distribution.

\begin{figure}
\plotone{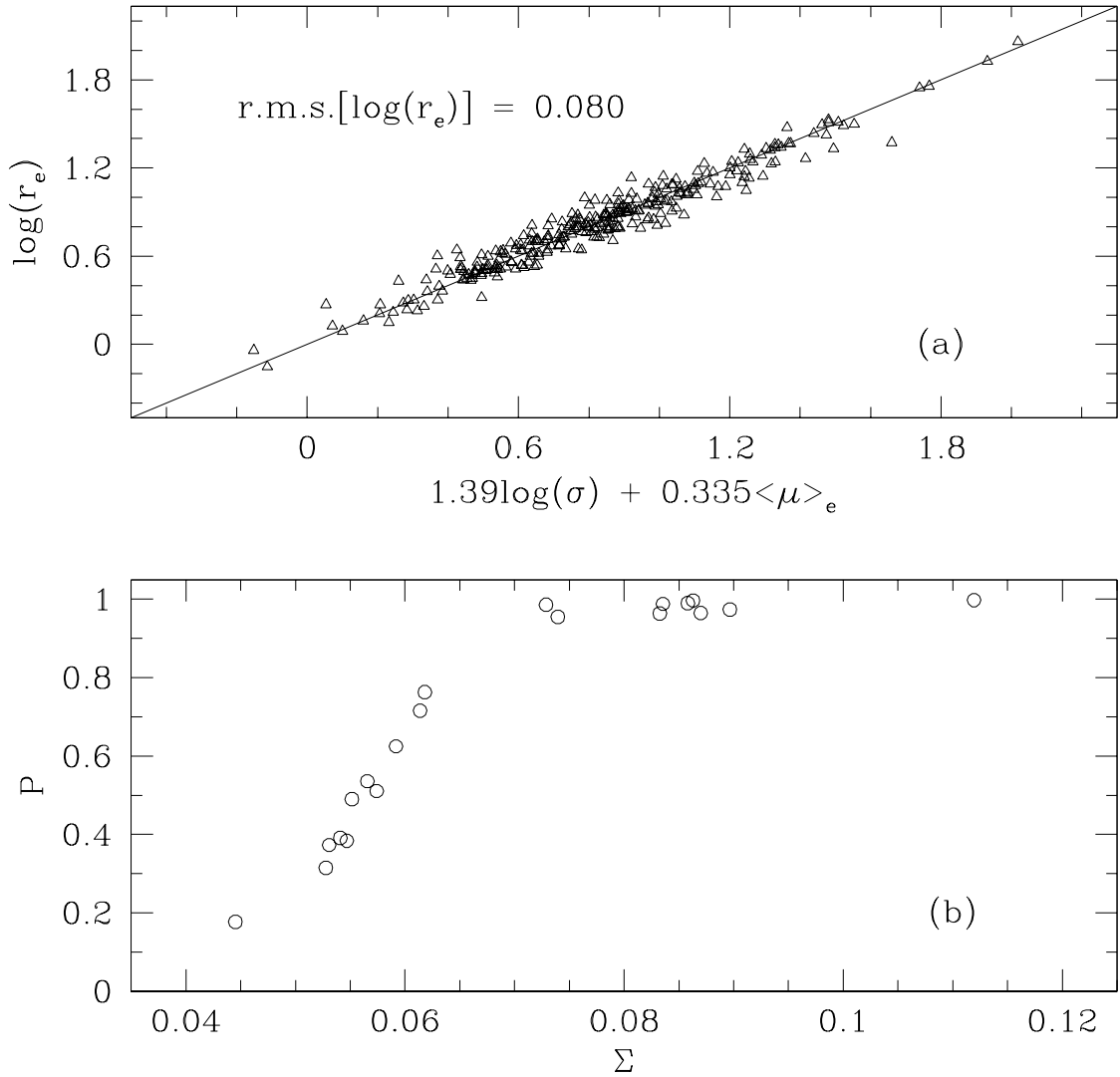}
\caption{\label{fig:fp11}(a) The \avgfp\ for the 11 best-fit clusters.
         (b) Same as Figure~\ref{fig:fp20}b but for the 11 cluster FP fit.
         When fit this way the high-\fpscat\ clusters are outliers, falling 
         in the upper $5\%$ tail in probability space, while the rest are 
         consistent with being uniformly distributed.}
\end{figure}

We therefore refit for \avgfp\ based on the 11 best-fit clusters 
(Figure~\ref{fig:fp11}a). Although the FP coefficients are only slightly 
different, $\alpha=1.39\pm$$0.04$, $\beta=0.335\pm$$0.005$, and \muscat\  
changes by $\sim10\%$, the 9 clusters with $\fpscat>\muscat$ clearly appear to 
be outliers, falling in the upper $5\%$ tail in probability space 
(Figure~\ref{fig:fp11}b). 
If truly deviant, these clusters may not be reliable objects for FP distance 
work. But do the high-\fpscat\ clusters distinguish themselves from the rest 
in any other way? PV is a second, independent dimension in which we can 
examine the behavior of this sample and is after all the quantity we aim to 
measure.

\subsection{\it Measuring Distances and Peculiar Velocities}
To derive peculiar velocities, we must first fix the scale between FP relative 
distances and absolute distances. That is, we must set the conversion 
between $\gamma$ and cluster redshift. We begin with the assumption that the 
clusters are at rest with respect to the Hubble flow, i.e. at rest in the 
CMB frame. Cluster line-of-sight velocities are translated 
from heliocentric redshifts, as determined from our spectra, to redshifts in 
the CMB frame using a vector in the direction $l=264\fdg 4$, $b=48\fdg 4$ in 
Galactic 
coordinates with a magnitude of 369.5~\kms, (Kogut \etal\ 1993).
Taking into account (the small) cosmological corrections, we then find the 
angular diameter $-$ distance zero-point which minimizes the
one-dimensional \rms\  peculiar velocity, \onedrms, of the clusters. This 
normalization does not change significantly if we minimize \onedrms\ for the 
eleven best-fit clusters instead of for all twenty. Interestingly, this 
procedure places the Coma cluster nearly at rest, so we would have 
recovered a similar result had we anchored the physical scale by assuming no 
peculiar motion for Coma (which has traditionally been done). 

\begin{figure}
\plotone{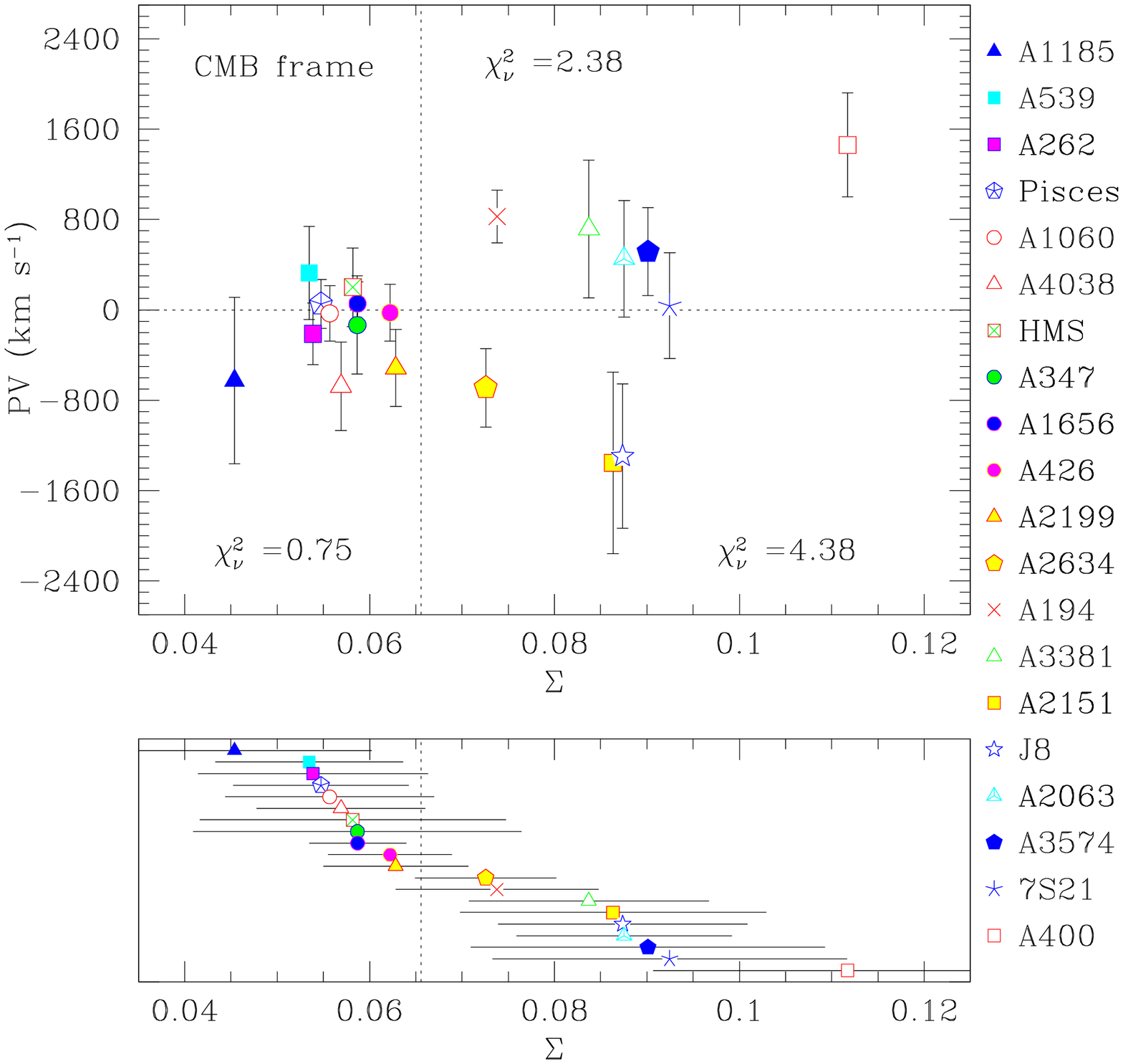}
\caption{\label{fig:pv}Cluster PV vs. \fpscat. Each point 
         represents a cluster containing $\sim10-80$ observed galaxies. The
         x-axis is the scatter within the cluster with respect to \avgfp. The
         abscissa is the PV as derived from redshift minus \avgfp\ distance. 
         The reduced $\chi^2$ about zero PV for the entire sample is shown at
         the top of the figure. \muscat\ (vertical line) roughly marks the 
         separation of the sample into two subgroups, over which $\chi^2$ 
         rises dramatically (values shown at the bottom of the panel). The 
         errors on \fpscat\ are shown in the bottom panel for clarity. One 
         can see a paucity of clusters about \muscat, where one would expect 
         the density of clusters to be highest if clusters are drawn from a
         single population.}
\end{figure}

\begin{figure}
\plotone{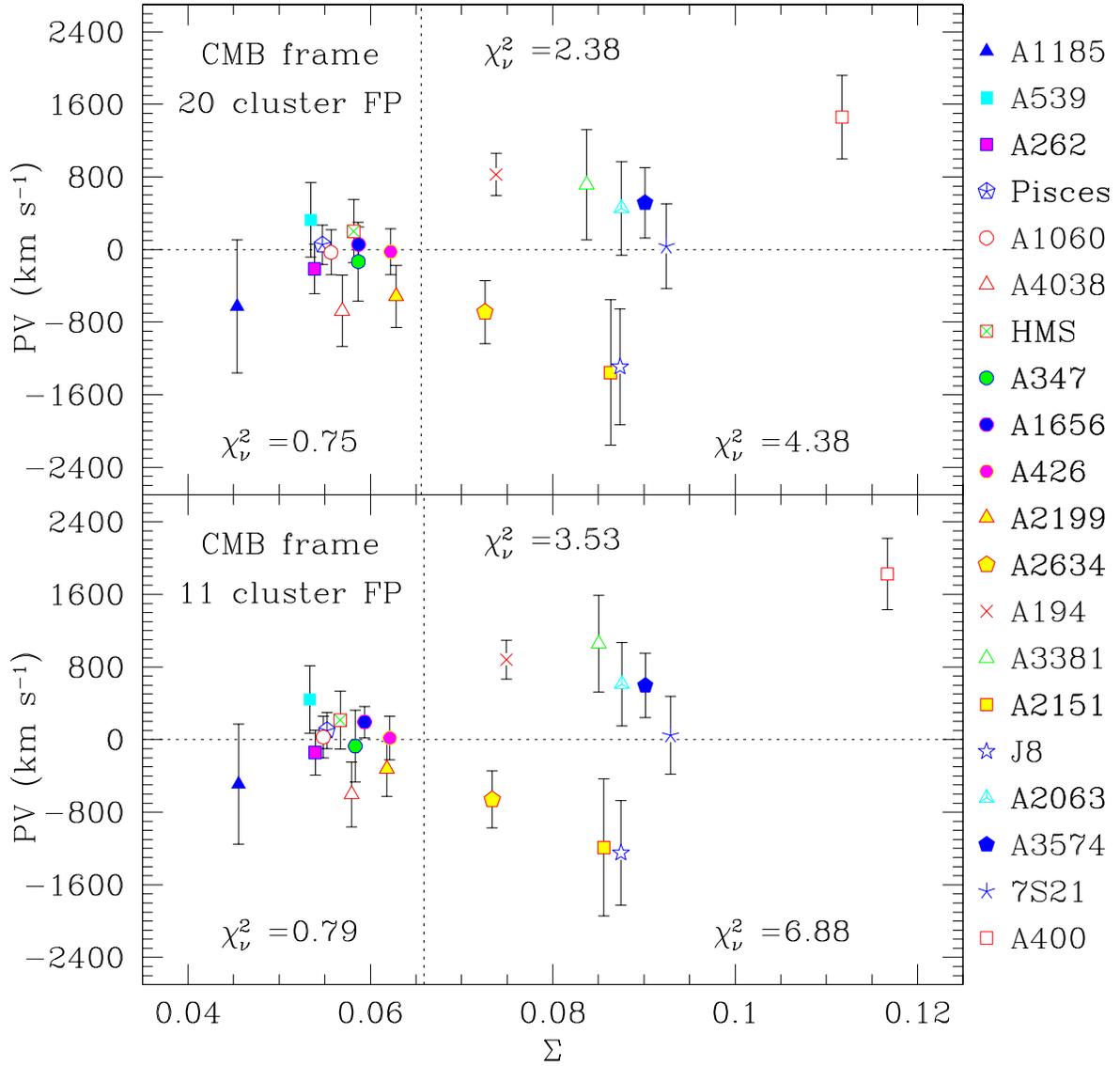}
\caption{\label{fig:20comp11}Comparison of the PV distribution for the 20 and
         11 cluster fits. As is readily seen, use of one fit or the other 
         does not significantly change our results.}
\end{figure}

Finally, plotting PV against \fpscat, we uncover an astonishing result 
(Figure~\ref{fig:pv}). There is an obvious discontinuity in the scatter in 
cluster PV which occurs across the same region of \fpscat-space as the gap 
in probability already discussed. The clusters which fit \avgfp\ well are 
at rest with respect to the CMB with reduced $\chi^2<1$ while the poor-fit 
clusters with $\fpscat>\muscat$ show large scatter in their PVs, $\chi^2=4.4$. 
\footnote{
Fractional errors in distance are predicted based on \muscat\ and the FP 
coefficient, $\alpha$, and scale inversely with $\sqrt{n_{gal}}$. Estimated PV 
errors are distance errors added in quadrature with redshift errors.
} 
Results from both the 20 cluster FP and 11 cluster FP are similar as can be 
seen in Figure~\ref{fig:20comp11}.

While one might naively expect a correlation between \fpscat\ and the 
measurement error of PV, this is not the case. If cluster ellipticals do indeed
form one population, then the sample mean and sample standard deviation of a 
randomly drawn subset will be uncorrelated. Therefore, PV should not reflect 
the value of \fpscat. Similarly PV {\it errors} do not correlate with \fpscat;
we therefore scale the errors by \muscat. As we will show in the 
following section, this null hypothesis is heavily disfavored as the observed 
non-uniformity in \onedrms\ is highly significant.

\subsection{\it Significant Variation in \onedrms} 
To examine the significance of the variation in \onedrms, we have used two 
approaches, linear regression and analysis of variance (ANOVA). Simple 
regression favors a strong correlation between PV and \fpscat whether we use 
\muscat\ to estimate the PV errors (null hypothesis), or use the individual 
\fpscat. Straightforward weighted regressions of ``$|$PV$|$ on \fpscat'' and 
``\fpscat\ on $|$PV$|$'' yield slopes which are nonzero with $99.9\%$ 
confidence. The true slope lies somewhere between the forward and inverse 
regressions. But because of the two-way errors, it is difficult to decide 
significance analytically, so the problem can also be dealt with numerically.
Resampling by complete permutation along the \fpscat\ axis is a measure of 
the likelihood of the observed correlation.
Permutation ignores errors in \fpscat\ and complete mixing of the 20 points
without replacement is allowed. Under the assumption of no correlation, the 
expected slope is zero, negative and positive slopes being equally likely. 
The resultant distribution shows the observed slope to be unlikely, with a 
probability of less than one percent.
\footnote{
Note, however, if two separate populations do not show such a correlation, this
test is inappropriate, in which case ANOVA is more meaningful.
}

The null hypothesis likewise fails ANOVA tests. The measured \onedrms\ for the 
two subsamples of clusters (defined by the gap in \fpscat) are not consistent 
with a single PV distribution. The ratio of chi-squares disfavors the two
groups being drawn from the same distribution with probability $99.94\%$.
\footnote{
This is not the traditional F-test because the comparison is not between two
independent fits. For this problem, the number of degrees of freedom for the 
subgroups is $\frac{nclust-1}{nclust}\times nsubgroup$. Specifically for 
$nclust=20$, the d.o.f. are $8.55$ and $10.45$ for $nsubgroup = 9$ and $11$ 
respectively. Furthermore, the ratio of $\chi^2$ is no longer purely 
F-distributed, so we have computed the proper cumulative distribution 
function.
}
Again permuting the points along \fpscat$-$space allows us to estimate the 
probability of the present result without an assumption about the nature of 
the distribution.
Using this bootstrapping procedure one finds the significance 
for the ratio of observed chi-squares is again above the $99^{th}$ percentile. 
The behavior of the sample both in \fpscat$-$space as well as in PV$-$space
tends to contradict at very high levels of significance the assumption
that we are observing a single population of clusters. Taken together, these 
statistics are convincing evidence that we are seeing two populations.
If this bimodality is an intrinsic property of clusters of galaxies, then all 
current FP samples potentially mix together these two apparently different 
populations.

\section{\label{Discuss}Errors in FP Distances or Peculiar Velocities?}
We have established that our sample of clusters divides into two groups 
defined by significantly different intrinsic scatter about the FP. 
Additionally, we find that these two groups have radically different PV 
distributions. The observation that only the high-\fpscat\ clusters show 
peculiar motions implies that such clusters can inject significant error 
into the measure of \onedrms. This could have the consequence of producing a 
spuriously large PV signal for the entire sample. Alternatively, 
these two groups could indeed be subject to truly different kinematics.
Either scenario, suggests that the nature of the inferred PV field is sample 
dependent.

We find, however, no evidence that this apparent separation of the clusters 
into two populations is due to systematic problems with our data. There is no 
correlation of \fpscat\ with cluster distance (e.g. more distant clusters
do not have poorer fits).
We see no effect due to data quality or data source; the clusters which 
we observed ourselves and the sample which was culled from the literature are 
equally divided between the two \fpscat\ groups, and with the exception of 
A400, the quality of the imaging and spectra of all the clusters are 
comparable. We are also confident that our cluster samples are not 
significantly contaminated by field ellipticals or S0s; we are sampling the 
centers of clusters over an area of approximately one half square degree and 
our redshift criterion is $\vert$z$-$z$_{mean}\vert\lesssim3\sigma$, where 
$\sigma=$ cluster velocity dispersion. We calculate that, with the possible 
exception of 7S21, the expected number of field interlopers per cluster in 
our sample is far less than one.
Addition of an Mg2 line index term to diminish cluster to cluster differences
in the age/metallicity of the stellar populations also does not improve
the fits. We find no strong evidence for curvature along the FP. In addition,
the galaxies in each cluster cover similar ranges in \logsigma\ so that we
expect any error introduced by sampling different portions of the FP is at 
most a minor effect in our final fits and distance measurements. The 
high-\fpscat\ clusters do on average have a smaller number of observed members
than low-\fpscat\ clusters. However, we would expect a small sample size to
only increase the scatter of \fpscat\ and not bias its value. Bootstrap tests
on subsamples of galaxies chosen at random from Coma show this to be the case.
Rather, the high-\fpscat\ clusters are also on average the least rich clusters,
and as we discuss below, there is good reason to believe that \fpscat\ is a 
function of cluster richness.

\begin{figure}
\plotone{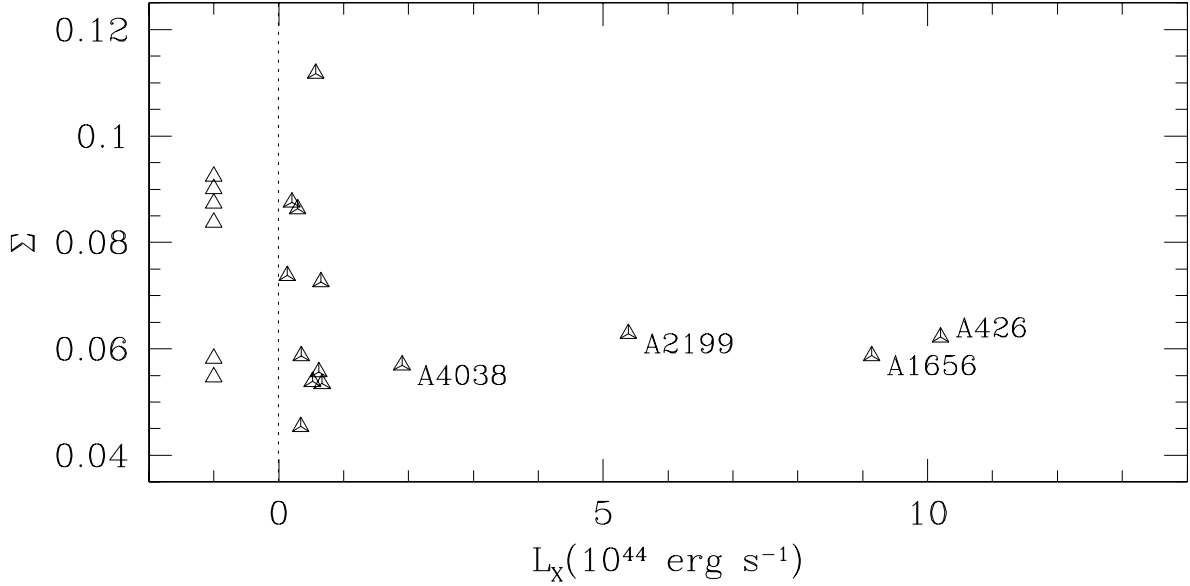}
\caption{\label{fig:xray}\fpscat\ vs. X-ray luminosity, L$_{\mbox{x}}$. Note 
         that the massive, virialized clusters fit the FP well, while the 
         quality-of-fit for the lower mass clusters cannot be predicted. The 
         most X-ray bright clusters (L$_{\mbox{x}}\gtrsim10^{44}$ erg s$^{-1}$)
         are labeled. The clusters for which there exist no X-ray data are 
         shown along the left side of the panel. L$_{\mbox{x}}$ from 
         White, Jones, \& Forman (1997).}
\end{figure}

If our study is relatively free from systematic errors, then our statistical
analysis shows that different clusters have different amounts of intrinsic
\logsigma\ scatter about the FP. But what property of a cluster might be 
driving this observed difference?  We have looked at many cluster properties as
predictors of whether a cluster will be a high- or low-\fpscat\ cluster and
have found only one strong indicator : cluster X-ray luminosity, as can
clearly be seen in Figure~\ref{fig:xray}. The four brightest X-ray clusters, 
those with L$_{\mbox{x}} > 10^{44}$ erg s$^{-1}$, all have low \fpscat. 
This result 
suggests that early-type galaxies within the most massive, well-virialized 
clusters form a more homogeneous population than cluster ellipticals as a 
whole.

\begin{figure}
\plotone{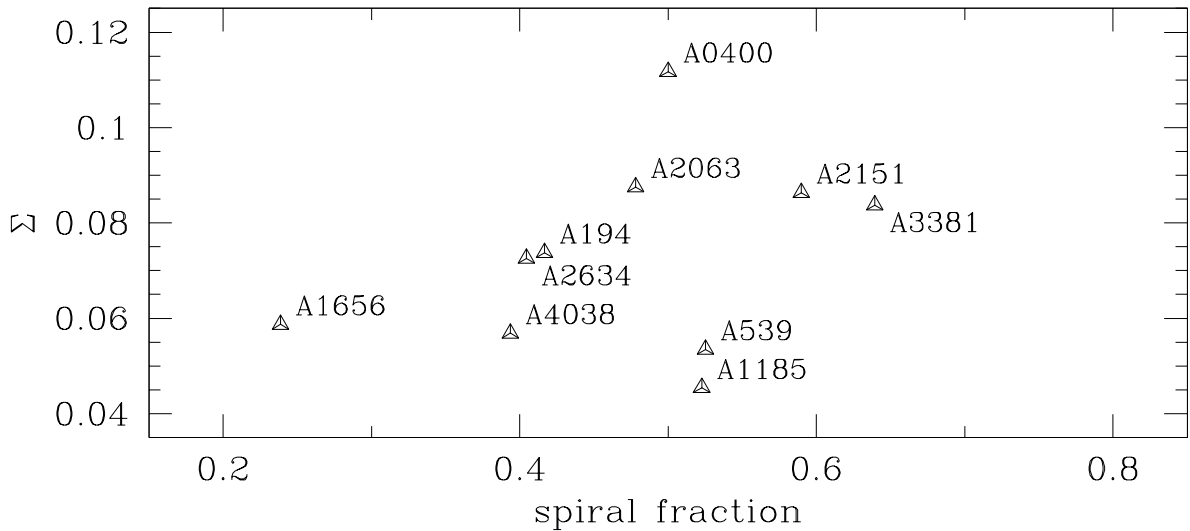}
\caption{\label{fig:sf}\fpscat\ vs. cluster spiral fraction (SF). A weak 
         correlation between these two variables is evident and is in the same
         sense as expected from the X-ray data. Spiral fractions from 
         Dressler 1980.}
\end{figure}

\begin{figure}
\plotone{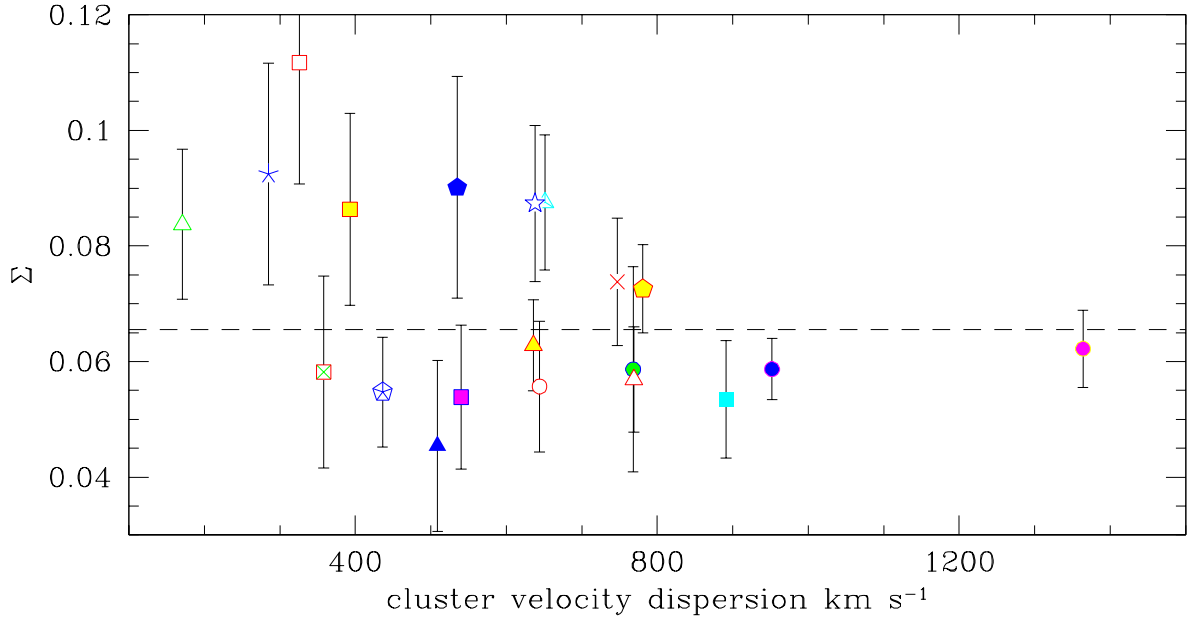}
\caption{\label{fig:vdisp}\fpscat\ vs. cluster velocity dispersion. The 
         high dispersion ($\gtrsim 800\ \kms$) clusters have low \fpscat.}
\end{figure}

A weaker, but similar correlation, can be seen with cluster spiral fraction, 
which is also an indirect indicator of the dynamical evolutionary state of a 
cluster (see for example Dressler \etal\ 1997). Clusters with $\fpscat > 0.08$ 
exhibit a spiral fraction of $55 \pm 1 \%$ compared to $42 \pm 3 \%$.
This trend of \fpscat\ with spiral fraction is shown in Figure~\ref{fig:sf}.
Similarly, \fpscat\ appears to be related to the cluster velocity dispersion, 
which is itself, of course, closely tied to cluster X-ray luminosity 
(Figure~\ref{fig:vdisp}). With one exception, clusters with velocity 
dispersions less than $400~\kms$ tend to have high \fpscat\ whereas those with 
dispersions greater than $800~\kms$ tend to have low \fpscat. 

There are also indications that substructure may be an important
component of the increased scatter observed in some clusters. Take the case of 
the relatively spiral rich, low velocity dispersion clusters A2151 and A400. 
In A2151, the early type galaxies in the core have a mean velocity offset of 
$-900~\kms$ from that of the spirals (Zabludoff \etal\ 1993, Dale \etal\ 1999b;
this paper). Indeed, A2151 and A2147 are clearly entangled in a mutual 
interaction (e.g. Bird \etal\ 1995; Maccagni \etal\ 1995), with a component 
of peculiar velocity reflecting infall and/or merging of subclusters. A 
similar offset between mean elliptical and cluster velocity has been observed
in A400 by Beers \etal\ (1992). In addition, both A400 and A2151 contain many 
examples of interacting/peculiar galaxies (see Bothun \& Schommer 1982),
indicating the prevalence of low velocity dispersion subgroups within the 
overall structure. Furthermore, substructures that are separated by $\sim 10$ 
Mpc or less will not be adequately resolved in kinematic distance space by the 
current relative indicators. An increase in \fpscat\ can be introduced due to 
a range in distance along the line of site. Indeed, such a spread in distances 
has already been proposed for one rich, but high \fpscat\ cluster, A2634 
(see West \& Bothun 1990; Scodeggio \etal\ 1995).

One reason for the fact that highly X-ray luminous clusters do not show 
significant peculiar velocity may be that X-ray luminosity coincides with the 
virialization of infalling structures. Whatever the source of the consistent 
behavior of these clusters, it appears that X-ray luminosity provides a good 
pre-filter for the selection of clusters to be used in PV studies. As will be 
shown in paper II of this series (Gibbons, Fruchter, \& Bothun 2000b), such 
a pre-filter produces a sample that exhibits no large scale flow.

\section{Conclusions}
 
Using high-quality data and a robust fitting technique, we have determined
that clusters of galaxies show a large range in quality-of-fit of their member
galaxies to the FP. Statistical tests described in \S4 argue that the
observed separation in \fpscat\ is significant at the $\sim 10^{-3}$ level.
Additionally, an examination of the distribution of clusters along the PV axis
provides further direct support for this separation. If clusters are truly 
random realizations of galaxies drawn from a single population, then the
total sample should have, on average, a uniform \onedrms. Instead, when
the sample is divided on the basis of the observed scatter about the FP,
the two subsamples have drastically different properties in the
independent dimension provided by the PV. Again, this effect is
statistically strong at more than the $99\%$ level. Simply put, high-\fpscat\
clusters have large PV and low-\fpscat\ clusters have essentially no PV, 
within the observational errors. Taken together, these two results cannot be 
reconciled with the idea that these clusters are randomly drawn from a single
population.

The behavior of \fpscat\ appears to be an intrinsic cluster property. We have 
found that a subset of the clusters, the X-ray luminous clusters, exhibit a 
well-defined FP. This suggests that the most massive and virialized clusters 
possess a homogeneous population of ellipticals and therefore provide the most 
accurate measures of relative distance. About half of the cluster sample 
possess \fpscat\ below the mean. When using this half of the sample, we find 
the clusters to be at rest within the CMB frame (see details in the 
forthcoming Gibbons \etal\ 2000b). The high-\fpscat\ clusters provide the bulk 
of the positive PV signal when all the clusters are averaged together.

Overall, our results strongly suggest that making a cut on the r.m.s. scatter
about the FP, or more strictly on cluster X-ray luminosity, will lead to a 
more reliable sample from which bulk flow properties can be established. The 
current level of disagreement between various observations of large scale 
flows (e.g. Table~\ref{tab:surveys}) may possibly reflect the fact that less 
reliable samples have been used.

\eject
\singlespace

\begin{table}
  \begin{center}
  \begin{tabular}{lcccc} \\
    Survey               &Method &PV w.r.t. CMB (\kms)  &Survey Depth (\kms) 
                                                        &\onedrms\ (\kms)\\
    ACIF                  &BCG    &$689\pm178\;(l=343\;b= 52)$ &15,000       
                                                        &$\sim 400$\\
    SMAC                  &FP     &$630\pm200\;(l=260\;b= -1)$ &12,000       
                                                        &$\sim 350$\\
    LP10K                 &IRTF   &$720\pm280\;(l=266\;b= 19)$ &12,000       
                                                        &$\sim 415$\\
    Dale \etal\ 1999 a\&b &IRTF  &$<200$                      &18,000        
                                                        &$341\pm93$\\
    Watkins 1997          &IRTF   &--                          &12,000       
                                                        &$265^{+106}_{-75}$\\
  \end{tabular}
  \caption{\label{tab:surveys}Recent Surveys : These surveys disagree in 
           both the magnitude and the direction of their derived flows.
           References appear in the text.
           {\it N.b.} The Watkins analysis is of subsets of the 
           Giovanelli \etal\ (1997) and Willick \etal\ (1995) surveys.}
  \end{center}
\end{table}

\begin{table}
  \begin{center}
  \begin{tabular}{l|| r@{.}l r@{.}l c c c c c c c c}
  \hline
  \multicolumn{1}{l||}{}                   
  &\multicolumn{7}{c}{}                    
  &Data                                    
  &\fpscat                                 
  &\multicolumn{2}{c}{}                    
  &\fpscat                              \\ 
  cluster                                  
  &\multicolumn{2}{c}{$l$(deg)}            
  &\multicolumn{2}{c}{$b$(deg)}            
  &cz(km/s)                                
  &$\sigma_{clust}$                        
  &n$_{gal}$                               
  &Ref.                                    
  &\avgfp                                  
  &$\alpha_{IND}$                          
  &$\beta_{IND}$                           
  &FP$_{IND}$              \\ \hline\hline 
  A1656  & 57&6 & 88&0 & 7215 & 952 &79 &1,4 &0.059 &1.54 &0.305 &0.052\\
  A400   &170&2 &-44&9 & 6708 & 326 & 6 &1   &0.112 &0.86 &0.301 &0.149\\
  A1185  &203&1 & 67&8 & 9923 & 509 &10 &1   &0.045 &1.62 &0.267 &0.057\\
  A2063  & 12&9 & 49&7 &10406 & 651 &17 &1   &0.088 &1.88 &0.299 &0.096\\
  A2151  & 31&6 & 44&5 &10188 & 393 & 9 &1   &0.086 &1.50 &0.257 &0.070\\
  HMS0122&130&2 &-27&0 & 4636 & 358 & 9 &5   &0.058 &1.87 &0.293 &0.068\\
  J8     &150&3 &-34&4 & 9425 & 638 &13 &5   &0.087 &1.37 &0.340 &0.096\\
  Pisces &126&8 &-30&3 & 4714 & 436 &25 &5   &0.055 &1.10 &0.369 &0.047\\
  A2199  & 62&9 & 43&7 & 8947 & 636 &36 &4   &0.063 &1.32 &0.359 &0.059\\
  A262   &136&6 &-25&1 & 4417 & 540 &15 &1,5 &0.054 &1.67 &0.308 &0.061\\
  A2634  &103&5 &-33&7 & 9132 & 781 &38 &1,4 &0.073 &1.37 &0.311 &0.072\\
  A347   &140&7 &-18&1 & 5312 & 768 & 8 &5   &0.059 &1.26 &0.348 &0.061\\
  A426   &150&5 &-13&7 & 5139 &1364 &49 &1,5 &0.062 &1.45 &0.357 &0.065\\
  7S21   &113&8 &-40&0 & 5517 & 285 & 7 &5   &0.092 &1.94 &0.179 &0.130\\
  A194   &142&2 &-62&9 & 5122 & 747 &19 &3,2 &0.074 &1.04 &0.333 &0.069\\
  A539   &195&6 &-17&6 & 8615 & 891 &22 &3   &0.053 &1.46 &0.327 &0.055\\
  A3381  &240&3 &-22&7 &11471 & 171 &14 &3   &0.084 &1.72 &0.227 &0.076\\
  A3574  &317&4 & 31&0 & 4873 & 535 & 7 &3   &0.090 &1.49 &0.352 &0.115\\
  A4038  & 25&3 &-75&8 & 8473 & 769 &27 &3,2 &0.057 &1.36 &0.342 &0.061\\
  A1060  &269&6 & 26&5 & 3976 & 644 &18 &3,2 &0.056 &1.58 &0.347 &0.056\\\hline
  \multicolumn{12}{l}{1 = this work; 2 = Lucey \& Carter 1988 (LC88);} \\
  \multicolumn{12}{l}{3 = J{\o}rgensen, Franx \& Kj{\ae}rgaard 1995a,1995b 
                      (JFK95a,JFK95b);} \\
  \multicolumn{12}{l}{4 = Lucey, Guzm\'{a}n, Smith \& Carter 1997 (LGSC97);}\\
  \multicolumn{12}{l}{5 = Smith, Lucey, Hudson \& Steel 1997 (SLHS97)}
  \end{tabular}
  \caption{\label{tab:sample}The Cluster Sample : column (1) cluster name;\ 
           (2-3) cluster position in Galactic coordinates;\  
           (4) heliocentric redshift;\ (5) cluster velocity dispersion;\
           (6) number of observed galaxies;\ (7) data references as defined
           above;\ (7) scatter about the average FP;\ (8-9) FP parameters for
           the fits to individual clusters.}
  \end{center}
\end{table}

\begin{references}
\reference{} Aaronson, M., Bothun, G. D., Mould, J. R., Huchra, J. P., 
             Schommer, R. A., \& Cornell, M. E. 1986, ApJ, 302, 536
\reference{} Akritas, M. G. \& Bershady, M. A. 1996, ApJ, 470, 706
\reference{} Bahcall, N. A. \& Fan, X. 1998, ApJ, 504, 1
\reference{} Bird, C. M., Davis, D. S., \& Beers, T. C. 1995, AJ, 109, 920
\reference{} Bothun, G. D. \& Schommer, R. A. 1982, AJ, 87, 1368
\reference{} Borgani, S., da~Costa, L. N., Freudling, W., Giovanelli, R., 
             Haynes, M. P., Salzer, J., \& Wegner, G. 1997, ApJL, 482, 121
\reference{} Burstein, D. \& Heiles, C. 1984, ApJS, 54, 33
\reference{} Beers, T. C., Gebhardt, K., Huchra, J. P., Forman, W., 
             Jones, C., \& Bothun, G. D. 1992, ApJ, 400, 410
\reference{} Coleman, G. D., Wu, C-C, \& Weedman, D. W. 1980, ApJS, 43, 393
\reference{} Courteau, S., Faber, S. M., Dressler, A., \& Willick, J. A. 
             1993, ApJL, 412, 51
\reference{} Carvalho, de R. R. \& Djorgovski, S. 1989, ApJ Lett., 341, 37
\reference{} Carvalho, de R. R. \& Djorgovski, S. 1992, ApJ Lett., 389, 49
\reference{} Dale, D. A., Giovanelli, R., Haynes, M. P., Campusano, L. E., 
             Hardy, E., \& Borgani, S. 1999a, ApJL, 510, 11
\reference{} Dale, D. A., Giovanelli, R., Haynes, M. P., Campusano, L. E., 
             \& Hardy, E. 1999b, AJ, 118, 1489
\reference{} Djorgovski, S. \& Davis, M. 1987, ApJ, 313, 59
\reference{} Dressler, A. 1987, ApJ, 317, 1
\reference{} Dressler, A., Oemler~Jr., A., Couch, W. J., Smail, I., 
             Ellis, R. S., Barger, A., Butcher, H., Poggianti, B. M.,
             \& Sharples, R. M. 1997, ApJ, 490, 577
\reference{} Dressler, A., Lynden-Bell, D., Burstein, D., Davies, R. L., 
             Faber, S. M., Wegner, G., \& Terlevich, R. 1987, ApJ, 313, 42
\reference{} Franx, M., Illingworth, G., \& Heckman, T. 1989, ApJ, 344, 613
\reference{} Gibbons, R. A., Fruchter, A. S., \& Bothun, G. D. 2000b, in prep
\reference{} Gibbons, R. A., Fruchter, A. S., \& Bothun, G. D. 2000c, in prep
\reference{} Giovanelli, R., Haynes, M. P., Herter, R., Vogt, N. P., 
             Wegner, G., Salzer, J. J., da Costa, L. N., 
             \& Freundling, W. 1997, AJ, 113, 22
\reference{} Giovanelli, R., Haynes, M. P., Wegner, G., da Costa, L. N., 
             Freundling, W., \& Salzer, J. J. 1996, ApJ Lett., 464, 99
\reference{} Gregg, M. D. 1995, ApJ, 443, 527
\reference{} Guzm\'{a}n, R. \& Lucey, J. R. 1993, MNRAS, 263, 47
\reference{} Hudson, M. J., Lucey, J. R., Smith, R. J., \& Steel J. 1997, 
             MNRAS, 291, 461
\reference{} Hudson, M. J., Smith, R. J., Lucey, J. R., Schlegel, D. J.,
             \& Davies, R. L. 1999, ApJL, 512, 79
\reference{} Jacoby, G. H., Branch, D., Ciardullo, R., Davies, R. L.,
             Harris, W. E., Pierce, M. J., Pritchet, C. J., Tonry, J. L., 
             \& Welch, D. L. 1992, PASP, 104, 599
\reference{} J{\o}rgensen, I., Franx, M., \& Kj{\ae}rgaard, P. 1995a, 
             MNRAS, 273, 1097
\reference{} J{\o}rgensen, I., Franx, M., \& Kj{\ae}rgaard, P. 1995b, 
             MNRAS, 276, 1341
\reference{} J{\o}rgensen, I., Franx, M., \& Kj{\ae}rgaard, P. 1996, 
             MNRAS, 280, 167
\reference{} Kogut, A. \etal\ 1993, ApJ, 419, 1
\reference{} Kelson, D. D., Illingworth, G. D., van Dokkum, P. G.,
             \& Franx, M. 1999, astro-ph/9911065 
             (accepted for publication in the ApJ)
\reference{} Kirkpatrick, S., Gelatt, C. D., Vecchi, M. P. 1983, 
             Science, 220, 671
\reference{} Kriss, J. 1992, IRAF package {\it stsdas.contrib.redshift}
\reference{} Lauer, T. \& Postman, M. 1994, ApJ, 425, 418
\reference{} Lucey, J. R., Bower, R. G., \& Ellis, R. S. 1991, MNRAS, 249, 
             755
\reference{} Lucey, J. R. \& Carter, D. 1988, MNRAS, 235, 1177
\reference{} Lucey, J. R., Guzm\'{a}n, R., Smith, R. J., \& Carter, D. 1997, 
             MNRAS, 287, 899
\reference{} Lynden-Bell, D., Faber, S. M., Burstein, D., Davies, R. L., 
             Dressler, A., Terlevich, R. J., \& Wegner, G. 1988, ApJ, 326, 19
\reference{} Maccagni, D., Garilli, B., \& Tarenghi, M. 1995, AJ, 109, 465
\reference{} Metropolis, N., Rosenbluth, A., Rosenbluth, M., Teller, A.,
             \& Teller, E. 1953, J. Chem. Phys., 21, 1087
\reference{} Sargent, W. L. W., Schechter, P. L., Boksenberg, A.,
             \& Shortridge, K. 1977, ApJ, 212, 326
\reference{} Scodeggio, M., Giovanelli, R., \& Haynes, M. P. 1998, 
             AJ, 116, 2728
\reference{} Scodeggio, M., Solanes, J. M., Giovanelli, R., \& Haynes, M. 
             1995, ApJ, 444,41 
\reference{} Smith, R. J., Lucey, J. R., Hudson, M. J., \& Steel J. 1997, 
             MNRAS, 291, 461
\reference{} Strauss, M. A., Cen, R., Ostriker, J. P., Lauer, T. R., 
             \& Postman, M.  1995, ApJ, 444, 507 
\reference{} Tonry, J. L., \& Davis, M. 1979, AJ, 84, 1511
\reference{} Vanderbilt, D. \& Louie, S. G. 1984, J. Comp. Phys., 56, 259
\reference{} Watkins, R. 1997, MNRAS, 292, 59
\reference{} West, M. J. \& Bothun, G. D. 1990, ApJ, 350, 36
\reference{} Whitmore, B. C., Gilmore, D. M., \& Jones, C. 1993, ApJ, 407, 
             489
\reference{} White, D. A., Jones, C., \& Forman, W. 1997, MNRAS, 292, 419
\reference{} Willick, J. A., Courteau S., Faber, S. M., Burstein, D.,
             \& Dekel, A. 1995, ApJ, 446, 12
\reference{} Zabludoff, A. I., Geller, M. J., Huchra, J. P., 
             \& Vogeley, M. S. 1993, AJ, 106, 1273
\end{references}
\end{document}